\documentclass[a4paper,10pt,twoside]{cpc-hepnp}

\usepackage{multicol}
\usepackage{graphicx}
\usepackage{booktabs}
\usepackage{amssymb,bm,mathrsfs,bbm,amscd}
\usepackage[tbtags]{amsmath}
\usepackage{lastpage}
\usepackage{CJK}
\usepackage{verbatim}
\usepackage{enumitem}
\usepackage{mhchem}

\begin{document}
\begin{CJK*}{GBK}{song}


\footnotetext[0]{}

\title{Improvement of Main Drift Chamber Monte-Carlo tuning model at BESIII\thanks{
Supported by Ministry of Science and Technology of China(2009CB825200),
Joint Funds of National Natural Science Foundation of China(11079008, U1232201),
Natural Science Foundation of China(11275266, 11205184, 11205182, 11275210) and SRF for ROCS of SEM
 }}

\author{%
      ZHANG Rui$^{1;1)}$\email{zhangrui310@mails.ucas.ac.cn}%
\quad XIAO Dong$^{2}$
\quad AN Fenfen$^{2}$
\quad GAO Yuanning$^{3}$\\
\quad HE Kanglin$^{2}$
\quad JI Xiaobin$^{2}$
\quad JIN Shan$^{2}$
\quad LI Weidong$^{2}$\\
\quad LI Weiguo$^{2}$
\quad LIU Huaimin$^{2}$
\quad LIU Kai$^{3}$
\quad SHEN Xiaoyan$^{2}$\\
\quad WANG Yifang$^{2}$
\quad WU Linghui$^{2}$
\quad Xu Qingnian$^{1}$
\quad YUAN Ye$^{2;2)}$\email{yuany@ihep.ac.cn}\\
\quad ZHANG Yao$^{2}$
\quad ZHAO Guang$^{2}$
\quad ZHENG Yangheng$^{1}$\\
}
\maketitle

\address{%
$^1$ University of Chinese Academy of Sciences, Beijing 100049, Peoples Republic of China\\
$^2$ Institute of High Energy Physics, Chinese Academy of Sciences, Beijing 100049, Peoples Republic of China\\
$^3$ Tsinghua University, Beijing 100084, Peoples Republic of China\\
}

\begin{abstract}
Based on real data, a new parameterized model of the Main Drift Chamber response is proposed. In this model, we tune the ratio of good hits and the residual distribution separately.
By data quality checking, the difference between simulation and data in track reconstruction efficiency reduces from 1\% to 0.5\% averagely for pion in $J/\psi\rightarrow \pi^{+}\pi^{-}\pi^{0}$, and the momentum resolution agreement improves significantly for proton in $J/\psi\rightarrow p\bar{p}$.

\end{abstract}

\begin{keyword}
BESIII, main drift chamber, Monte-Carlo tuning, hit category
\end{keyword}

\begin{pacs}
29.40.Gx, 29.40.Cs, 13.66.Jn
\end{pacs}

\footnotetext[0]{\hspace*{-3mm}\raisebox{0.3ex}{$\scriptstyle\copyright$}2013
Chinese Physical Society and the Institute of High Energy Physics
of the Chinese Academy of Sciences and the Institute
of Modern Physics of the Chinese Academy of Sciences and IOP Publishing Ltd}%

\begin{multicols}{2}

\section{Introduction}\label{sec:introduction}
The Beijing Spectrometer \uppercase\expandafter{\romannumeral3} (BES\uppercase\expandafter{\romannumeral3})\cite{1}, which operates at the upgraded Beijing Electron Positron Collider (BEPC\uppercase\expandafter{\romannumeral2}), consists of the following sub-detectors: Main Drift Chamber (MDC), Time of Flight Counter (TOF), Electromagnetic Calorimeter (EMC), and Muon Counter (MUC). The MDC is the core sub-detector, measuring the decay vertex, energy loss and momentum of charged particles precisely. There are 6796 sense wires with a positive high voltage (HV), most of which are surrounded by 8 field wires. A drift cell is defined as the sensitive region of a sense wire. The whole MDC is in a 1.0~Tesla magnetic field provided by a super-conducting solenoid between EMC and MUC.

The MDC plays an important role in reconstructing charged tracks and identifying particles. A charged particle passing the drift cell ionizes the surrounding atoms of gas. The primary electrons drift towards the sense wire, being accelerated by electric field in the drift cell, and thereby initiating electron avalanches. These stripped electrons are eventually collected by sense wires and then produce electronic signals. If the magnitude of a signal exceeds a threshold value, relevant information will be recorded by the electronic system. This fired wire is called a hit.

Since the first physics data taking in 2009, the peak luminosity of BEPCII has reached $7\times 10^{32}~\textrm{cm}^{-2}\textrm{s}^{-1}$, which is about 70 times of BEPC. High luminosity and huge data sample reduce the statistical error of physical measurement significantly, and also call for a decrease of systematic uncertainty. A reliable Monte-Carlo (MC) simulation is important since event selection, efficiency calculation, and background estimation in the physics analysis, all rely heavily on it. Based on the GEANT4\cite{2} package, the BESIII Object Oriented Simulation Tool (BOOST)\cite{3} is developed on the framework of the BESIII Offline Software System (BOSS)\cite{4}. For physics analysis, good consistencies of tracking efficiency and momentum resolution between MC and data are crucial. In this letter, we describe a new method based on hits category to improve the MDC MC tuning (short for ``tuning'').

\section{Tuning principles}\label{sec:principles}

 The tracking efficiency is positively related to the raw hit efficiency, which is the probability of sense wire firing when a charged particle passes though the drift cell. The momentum resolution is mainly contributed by intrinsic resolution and multiple scattering. The latter factor has been considered in GEANT4 package. The intrinsic resolution of MDC reflects the errors of particle position measurement. It contains the effects of primary ionization position, electron diffusion along the drift path, amount and positions of avalanches, and distortion of the electric field at the edges of the drift cell. These processes are complicated for simulation from the first principle due to the limitation of computing power. The spatial resolution effect composes of intrinsic resolution contribution, multiple scattering, beam background, electronic noise, and tracking algorithm.

 Thus we build a parameterized model based on the raw hit efficiency and the spatial resolution for tuning. The values of parameters are extracted from the relevant distributions of real data. In order to get a better consistency, the iteration procedures are applied: using the model with initial values to generate simulation sample; then, comparing its relevant distributions with real data's and adjust the input values.

 In practise, the spatial resolution is described by residual distribution. Therefore, raw hit efficiency and residual distribution are two key indicators when tuning.

\section{Previous method}

With the past experience from BES\uppercase\expandafter{\romannumeral2}\cite{5}, we find it difficult to obtain a consistency of the two key indicators simultaneously, by only tuning residual distribution. A better strategy is to tune the raw hit efficiency and the residual distribution separately\cite{6}.

\subsection{Raw hit efficiency}
 The raw hit efficiency is given by:
 \begin{equation}\label{equ:hiteff}
  \epsilon_{H}=\frac{N_{hit}}{N_{pass}},
 \end{equation}
 where $N_{pass}$ is the number of drift cells passed by track; $N_{hit}$ means the number of drift cells fired in $N_{pass}$. We use helix parameters given by reconstruction algorithm to predict $N_{pass}$.

 The $\epsilon_{H}$ is a function of cell position, distance of closest approach (DOCA), and dip angle. So we divide it into three parts: $\epsilon_{cell}$, $\epsilon_{doca}$, and $\epsilon_{\cos{\theta}}$ and tune them separately.

\subsection{Residual distribution}
 The residual of a hit is defined as:
 \begin{equation}\label{equ:residual}
 r=d_{drift}-d_{doca},
 \end{equation}
 where $d_{drift}$ is the drift distance of the hit, calculated by drift time and distance-time relation (X-T relation); the DOCA, $d_{coca}$, is calculated by the parameters of the helix --- which are obtained from the other hits on this track.

 The residual distribution depends on layer, DOCA, and entrance angle. Thus, all hits are categorized by different layers, DOCA, and entrance angles, and we tune each residual distribution one by one. Double-Gaussian is used to describe residual shape of signal. And a 2-order Polynomial is employed to describe the shape of background and noise. In simulation, the contribution of beam background has been considered by random background sampling. Thus the polynomial part will not be tuned. As a result, there are five parameters for the residual distribution: ($f$, $\mu_{1}$, $\sigma_{1}$, $\mu_{2}$, $\sigma_{2}$), in which $f$ stands for the fraction of the first gaussian; $\mu_{1}, \mu_{2}$ stands for the mean values of the two gaussian; and $\sigma_{1}, \sigma_{2}$ stands for the standard deviations.

\subsection{Limitation and discussion}
 In the previous method, the average of tracking efficiency difference could be reduced to 1\%. Further comparison shows that a part of residual distributions near the sense wire has a tail at positive side, especially in inner layers.

 The residual tends to be positive due to its definition (Eq.~\ref{equ:residual}). As shown in Fig.~\ref{fig:asymmetry}, typically, $d_{doca}$ will not be larger than $d_{drift}$. This kind of discrepancy is enlarged in inner layers, where the lower working voltage than designed is applied to protect detector from high level of background. It requires more primary ionization to trigger the electronics. This could make X farther from C. As a result, the residual could be much larger.

\begin{center}
\includegraphics[width=3.5cm]{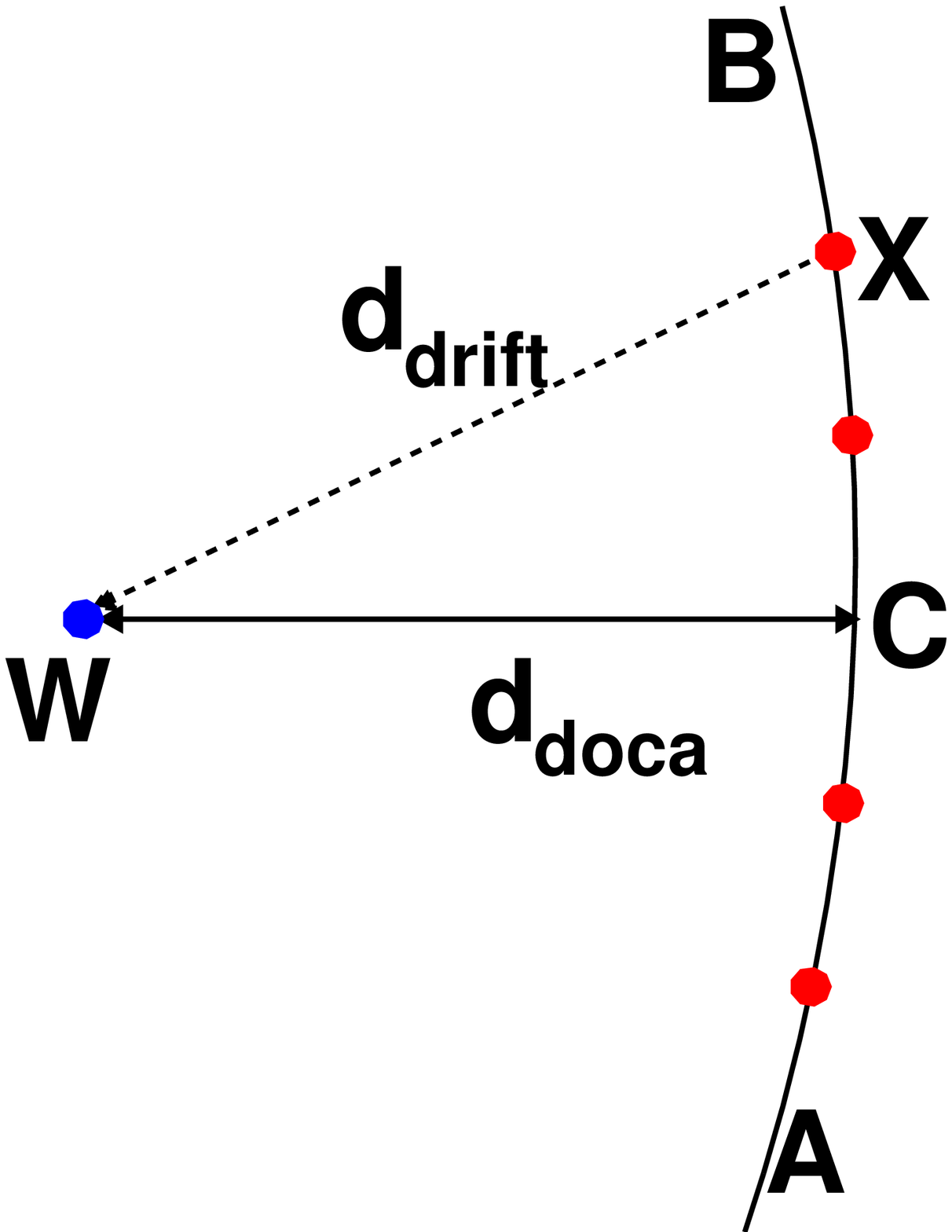}
\figcaption{\label{fig:asymmetry}   A schematic diagram of residual. W is a sense wire. A particle flies along $\widehat{AB}$. $\overline{WC}$ is DOCA. Solid dots on $\widehat{AB}$ are the positions of primary ionization. Suppose that X triggers the electronics. $\overline{WX}$ is always longer than $\overline{WC}$. }
\end{center}

 In order to verify the judgement, we select electron tracks from a Bhabha sample which is collected with full high voltage on inner layers. In Fig.~\ref{fig:longtail}, plot (a) is inner layer with normal voltage. Plot (b) is inner layer with full voltage. We also put an outer layer residual shape with full voltage as a reference (plot (c)). The conclusion is that with lower voltage on the wire, the tail tends to include more signal hits. And since these hits have no relationship with beam background, they can not be taken into account in neither previous tuning nor random background sampling.

\end{multicols}
\ruleup

\begin{minipage}{5.5cm}
\begin{center}
\includegraphics[width=5.6cm]{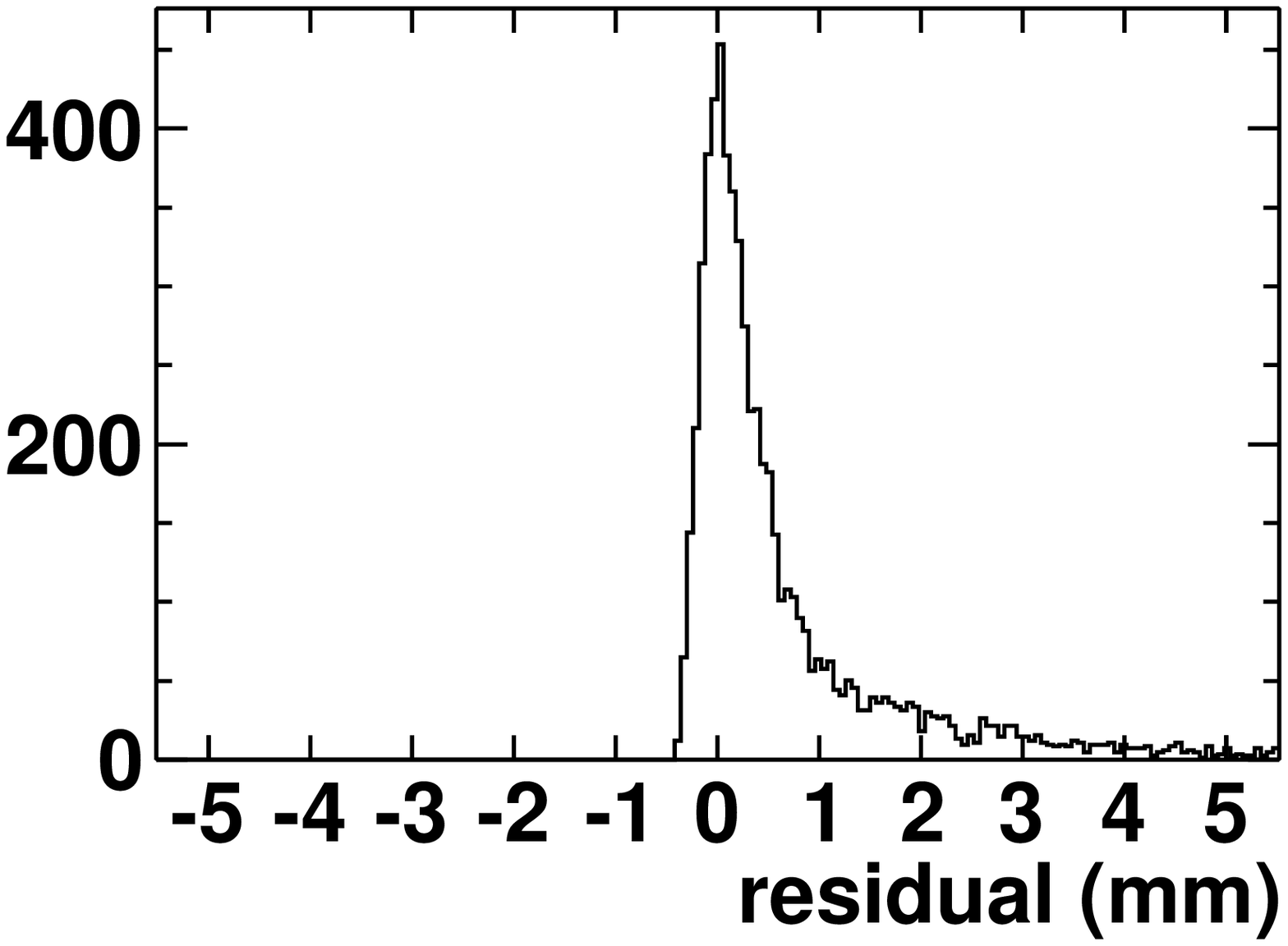}
\centerline{(a)}
\end{center}
\end{minipage}
\hfill
\begin{minipage}{5.5cm}
\begin{center}
\includegraphics[width=5.6cm]{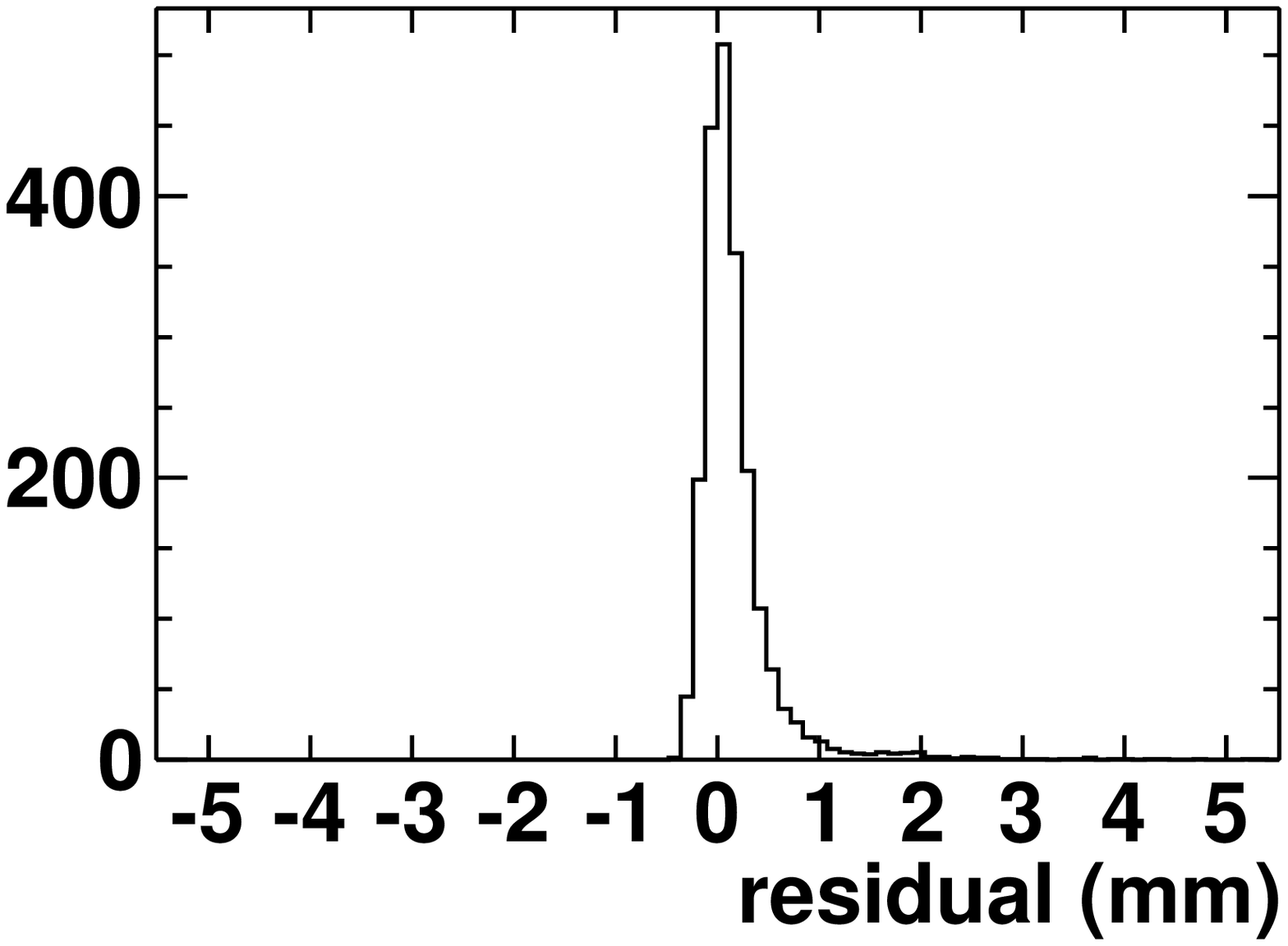}
\centerline{(b)}
\end{center}
\end{minipage}
\hfill
\begin{minipage}{5.5cm}
\begin{center}
\includegraphics[width=5.6cm]{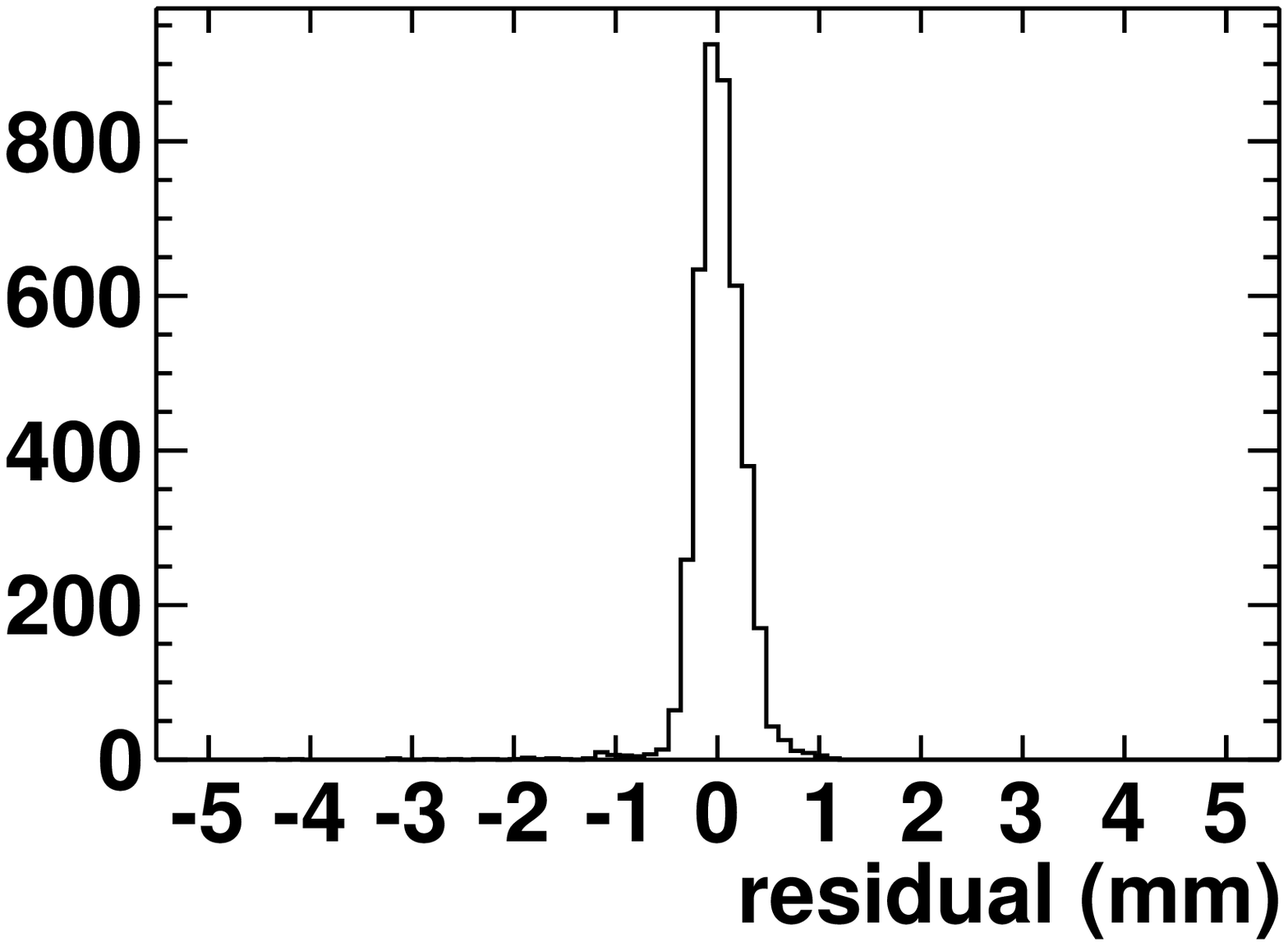}
\centerline{(c)}
\end{center}
\end{minipage}
\figcaption{\label{fig:longtail}   Residual distributions. Plot (a) is inner layer with lower HV --- the tail contains signals with large residual. (b) is inner layer with full HV --- the tail diminishes. (c) is outer layer with full HV. }
\ruledown

\begin{multicols}{2}

\section{New method}\label{sec:new method}
In order to improve the agreement between data and MC, a new method is developed to consider those signals with relatively larger residual.

\subsection{Hits category}
 The residual distribution is obtained by track reconstruction algorithm\cite{7,8}. The reconstruction algorithm drops those hits which make large contribution to $\chi^{2}$ when fitting. As a result, quality of a track is dominated by the hits adopted. In the previous method, we tune all hits as a whole. However, the consistency of the five parameters of double-Gaussian does not necessarily ensure the agreement of the residual resolution of the hits on track.

 A natural way of classification of all hits is based on whether a hit is used by tracking reconstruction: the hits used eventually on track are good hits while those not used on track are bad hits.

 The good-hit efficiency should be one of the criteria of the track quality. It is defined as:
 \begin{equation}
 \epsilon_{ghit}=\frac{N_{ghit}}{N_{pass}},
 \end{equation}
 where $N_{ghit}$ is the number of good hits; $N_{pass}$ is the same as in {Eq.~\ref{equ:hiteff}}. In the previous method, getting a better consistency of the good-hit efficiency requires to broaden or shrink the shape of double-Gaussian. This will change the good hits residual distribution. Thus it is difficult to reduce the disagreement between simulation and data further.

\subsection{Revised model}
Since the raw hit efficiency is tuned before residual distribution, we retain the model of the raw hit efficiency, but rebuild a model of the residual distribution. We use a parameter $Ratio$ to describe the proportion of good hits, a double-gaussian function with five parameters (two means, two standard errors and a fraction) to describe the residual distribution of good hits, and a uniform distribution of large residual to describe the shape of bad hits.

 Because a new parameter, $Ratio$, is introduced, double-Gaussian only focus on good hits distribution. This not only reduces the impact on good-hit efficiency, but also makes the fitting easer since the shape becomes better (Fig.~\ref{fig:newMethod}).

\subsection{Tuning procedures}
The process of tuning raw hit efficiency keeps the same with the previous method. After that, we tune the residual distribution. Due to the complicated correlation among the parameters, we make an order by their sensitivity. The $Ratio$ is tuned before the double-Gaussian shape.

\subsubsection{Obtaining the initial values}
 We get the initial values of the 6 parameters from the residual distribution of real data. The $Ratio$ is calculated by counting the numbers of good and total hits (Fig.~\ref{fig:newMethod} (a)). The rest five parameters are obtained by fitting the good hits distributions (Fig.~\ref{fig:newMethod} (b)).
\begin{minipage}{.45\linewidth}
\begin{center}
\includegraphics[width=4.3cm]{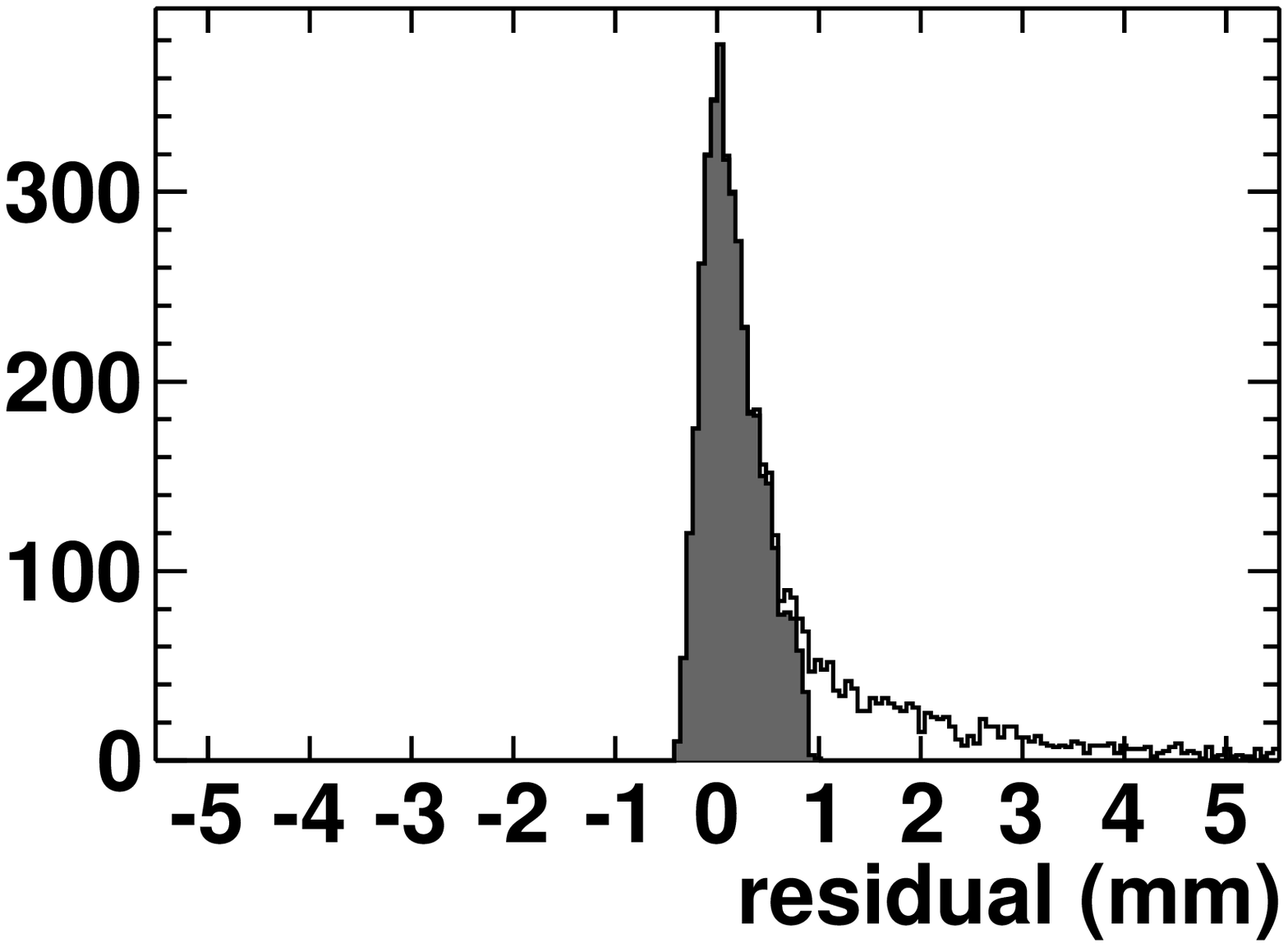}
\centerline{(a)}
\end{center}
\end{minipage}
\begin{minipage}{.45\linewidth}
\begin{center}
\includegraphics[width=4.3cm]{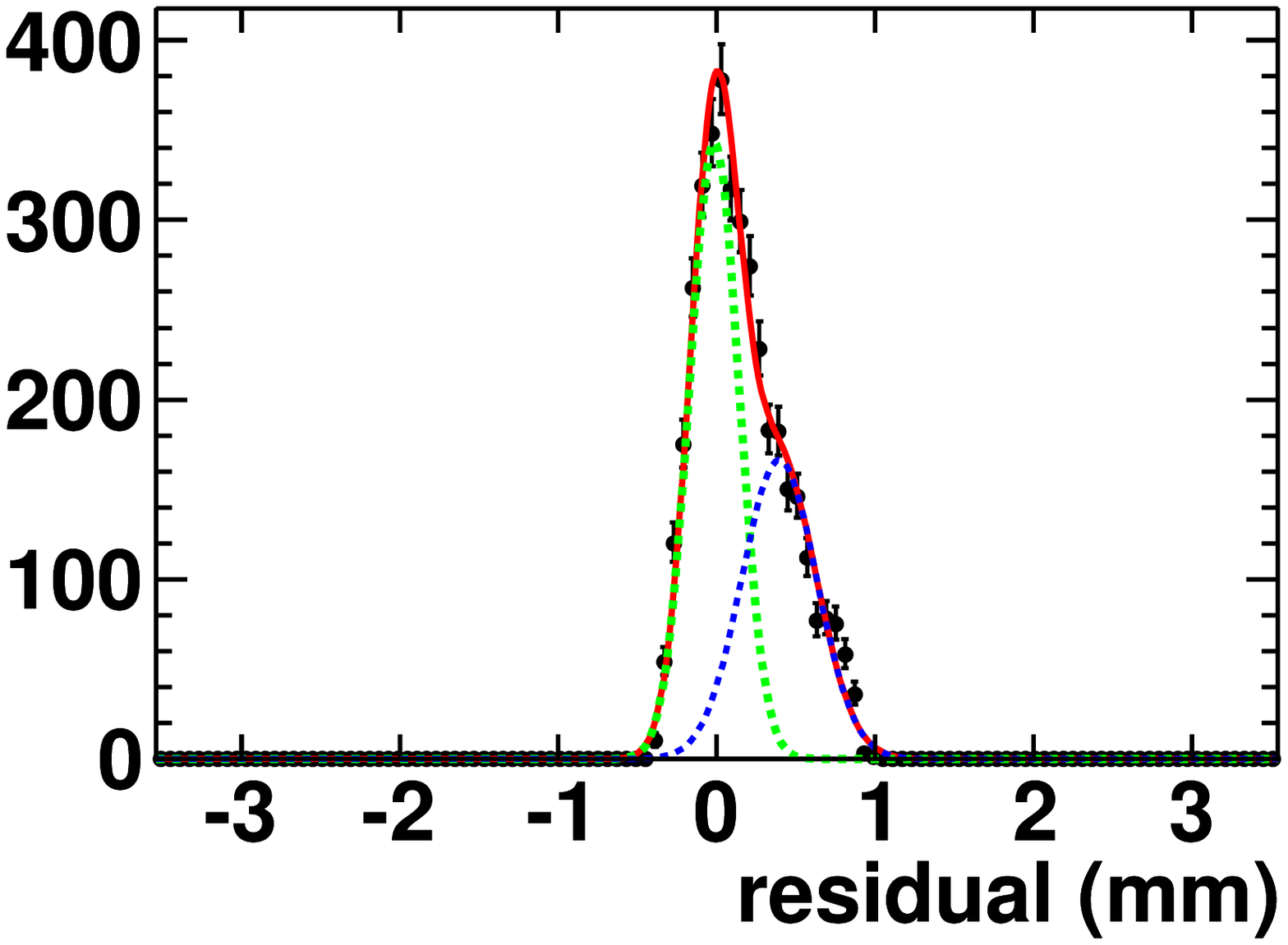}
\centerline{(b)}
\end{center}
\end{minipage}
\figcaption{\label{fig:newMethod}   (a) Two hit components: good hits (shaded) and bad hits (blank). (b) Fitting result of good hits distribution by double-Gaussian. }

\subsubsection{Ratio tuning}
We generate a MC sample by these initial values and compare the value of $Ratio$ with real data's. Based on the discrepancy, we change the input value while keep the other parameters fixed.

\subsubsection{Double-Gaussian identification}
 In order to decide which $\mu$ or $\sigma$ should be modified, it is necessary to judge which Gaussian describing the certain part of the shape. Since $\sigma{}$ is changing during tuning, we use $h$ as the identifier:
\begin{equation}
  h=f/\sigma,
\end{equation}
where $f$ is the fraction of the Gaussian; $\sigma$ is the standard error. The Gaussian with larger $h$ is labeled as $g_{1}$, and the other one is $g_{2}$.

\subsubsection{Gaussian shape tuning}
The most sensitive parameter of double-Gaussian is $\sigma_{1}$ since $g_{1}$ plays a main role in residual distribution. After several iterations, the main part of the residual shape keeps consistent. Then we tune $\sigma_{2}$, $f$, $\mu_{1}$, and $\mu_{2}$ in sequence.

\subsubsection{Tuning performance}
We compare the raw hit efficiency and good-hit efficiency with real data after tuning, respectively (Fig.~\ref{fig:performance}). After tuning, the efficiencies of MC and data agree well.

\begin{minipage}{4.2cm}
\begin{center}
\includegraphics[width=4.2cm]{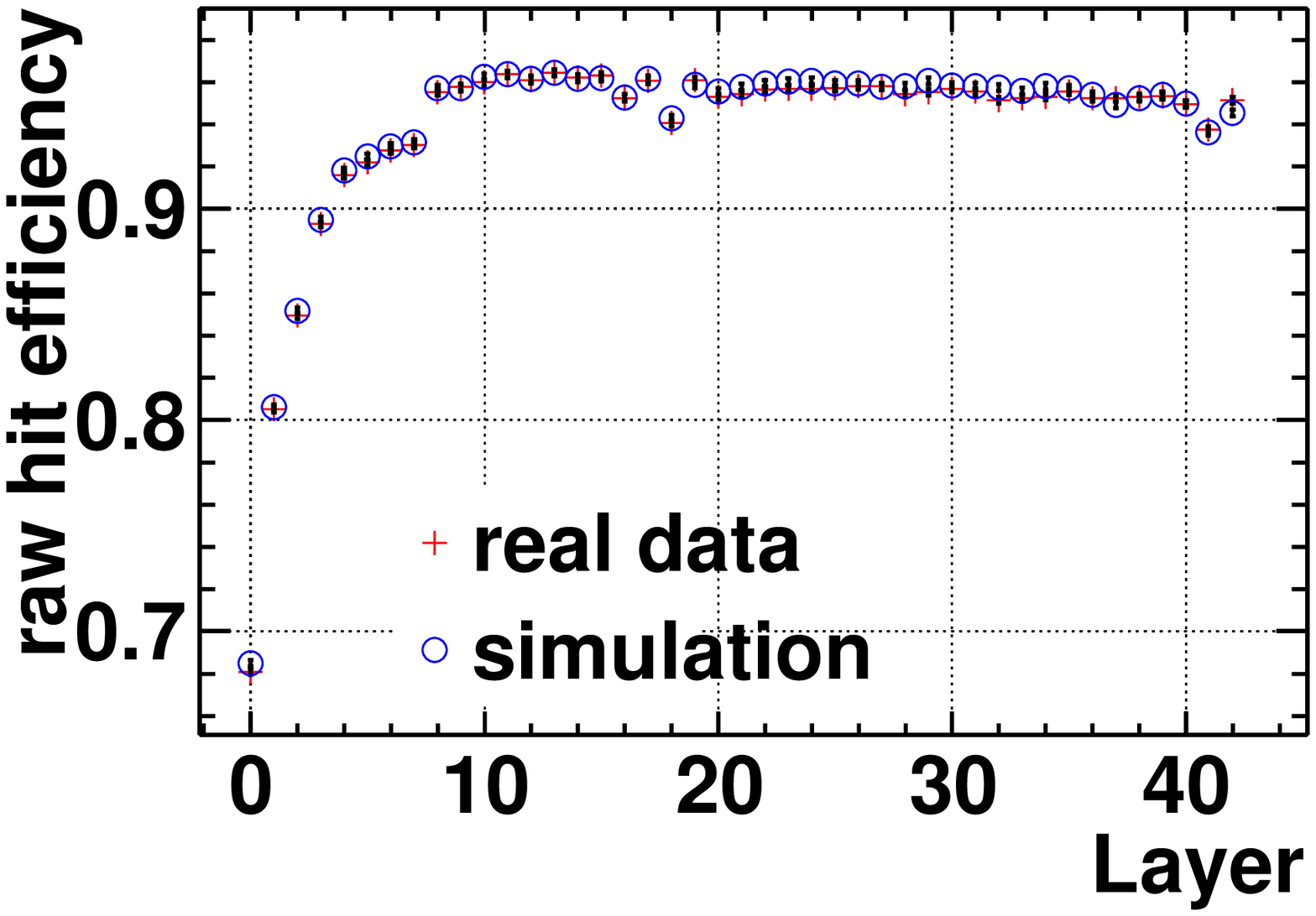}
\centerline{(a)}
\end{center}
\end{minipage}
\begin{minipage}{4.2cm}
\begin{center}
\includegraphics[width=4.2cm]{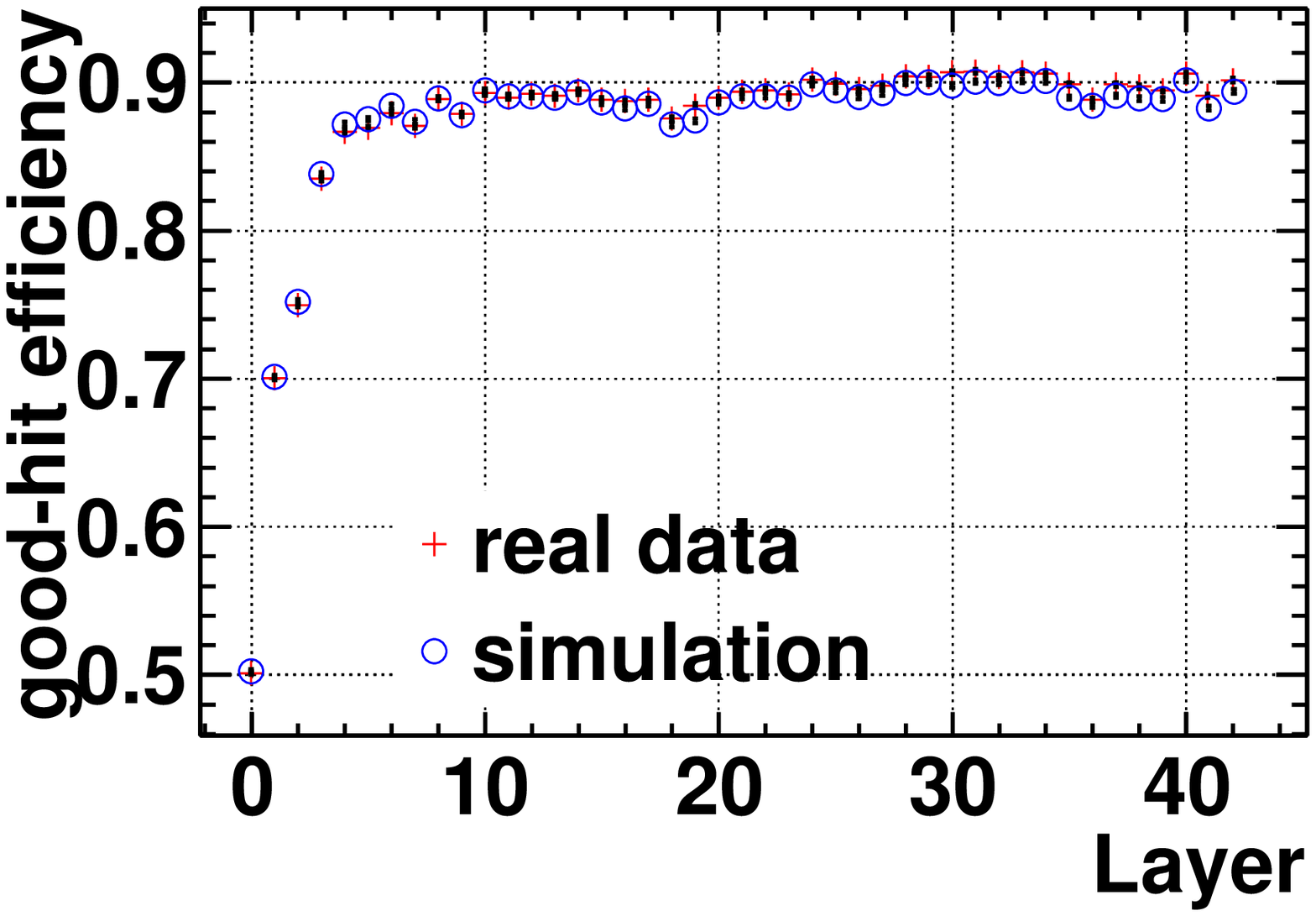}
\centerline{(b)}
\end{center}
\end{minipage}
\figcaption{\label{fig:performance}   Tuning results and comparisons to data. (a) raw hit efficiency in each layer, (b) good-hit efficiency in each layer. }

We also compare the difference of root mean square (RMS) of residual distribution in the two method in each layer (Fig.~\ref{fig:RMScmp}). After the improvement, the difference between MC and data become much smaller.

\begin{center}
\includegraphics[width=8cm]{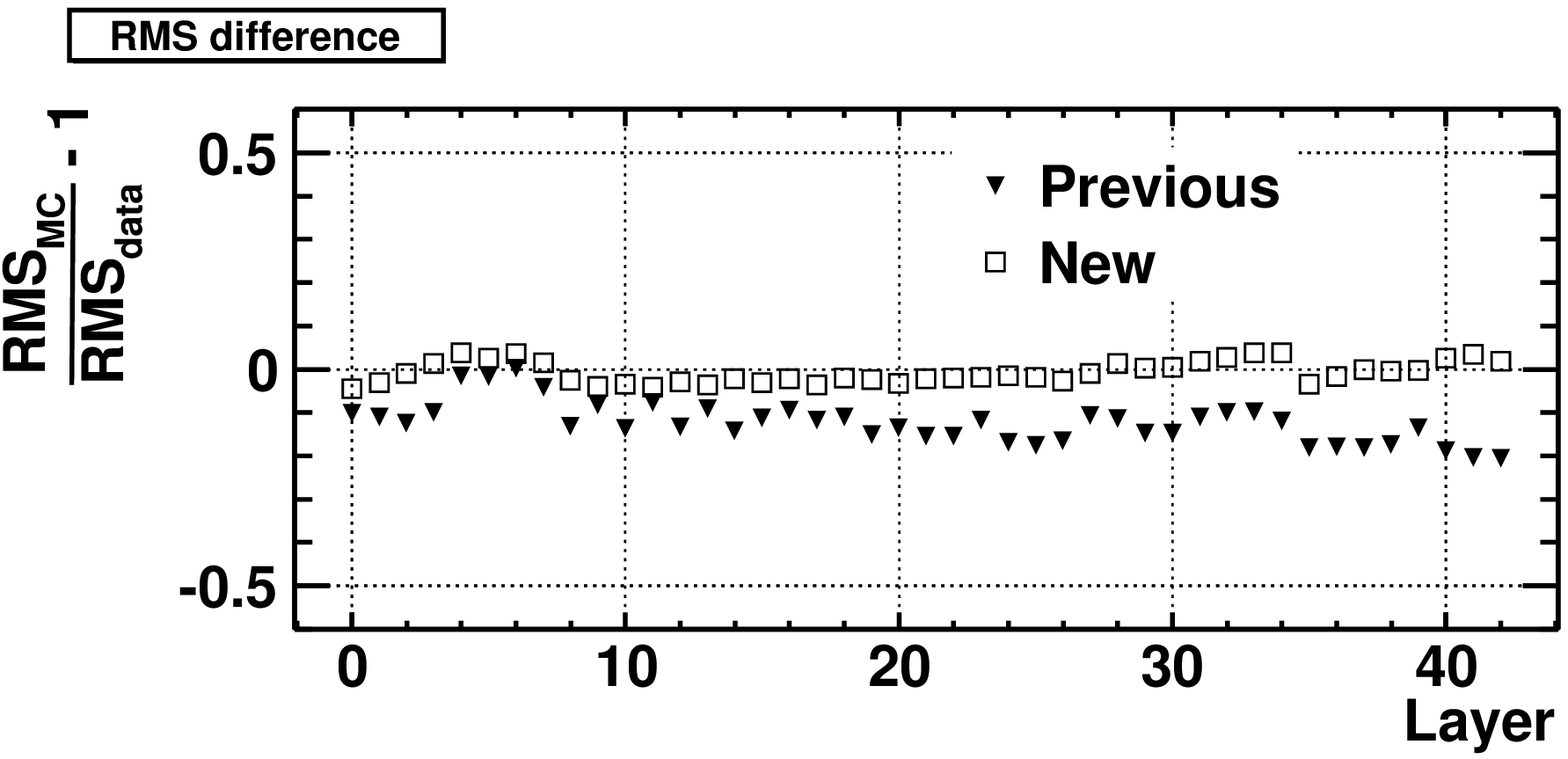}
\figcaption{\label{fig:RMScmp}   RMS difference between MC and data. Full triangles down stand for the previous method and open squares stand for the improved method.  }
\end{center}

\section{Check with physics objects}
In order to test the new method, we pick out some typical physics channels to compare results from simulation and real data.

\subsection{Tracking efficiency}
We use $\pi$ from $J/\psi\rightarrow \pi^{+}\pi^{-}\pi^{0}$ to check the difference of tracking efficiency. Tracking efficiency is defined as:
\begin{equation}
  \epsilon=\frac{N_{found}}{N_{recoi}},
\end{equation}
where $N_{recoi}$ is number of events in which $\pi^{0}$ and tagged charged pion have been reconstructed and the recoiled mass falls in the region of $\pi$; $N_{found}$ is the number of events from $N_{recoi}$ in which the recoiled $\pi$ is reconstructed successfully. Fig.~\ref{fig:pieff} shows that the difference of tracking efficiency between MC and data is about 0.5\% averagely.
\begin{minipage}{9cm}
\begin{center}
\includegraphics[width=7.9cm,height=4cm]{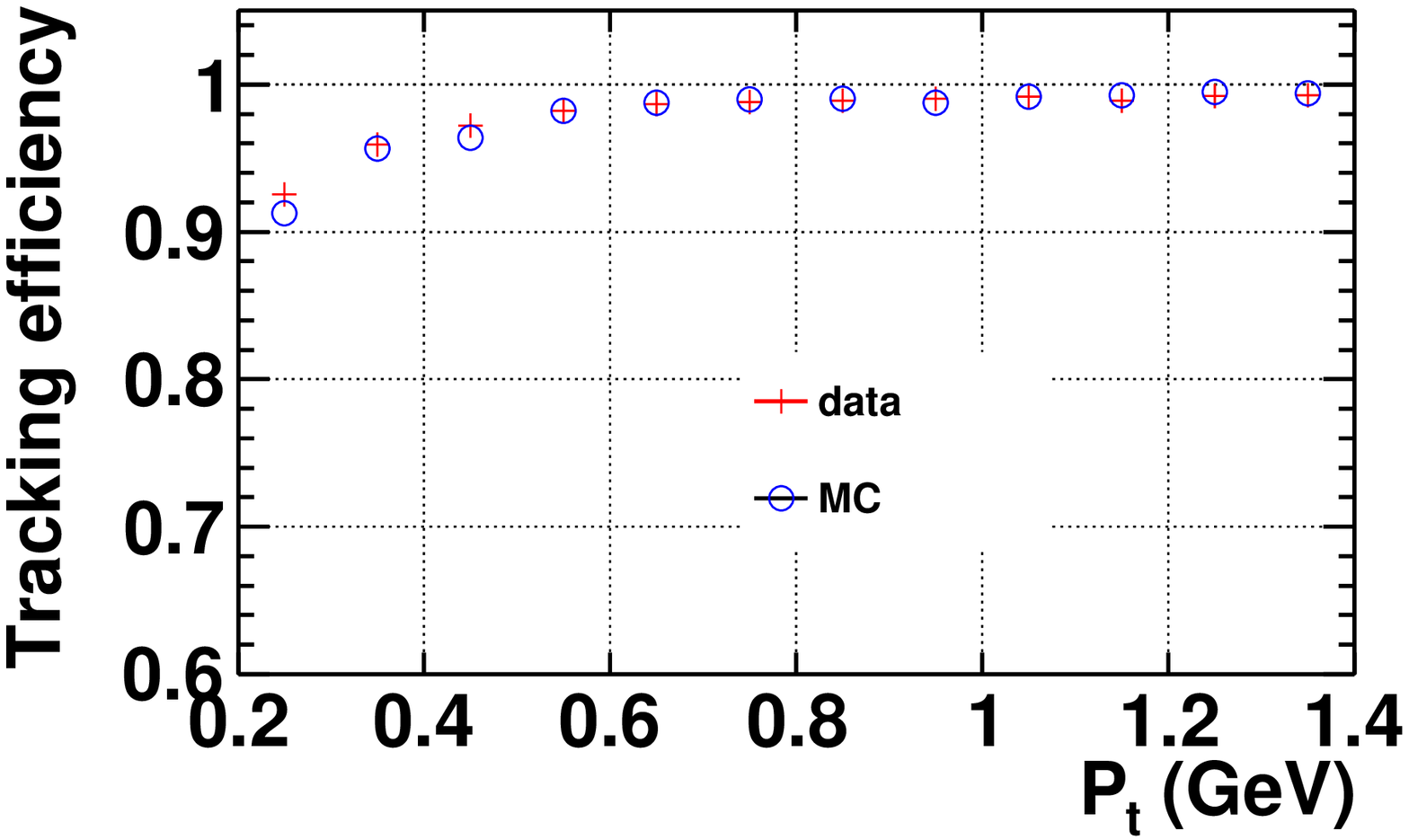}
\centerline{(a)}
\end{center}
\end{minipage}
\begin{minipage}{9cm}
\begin{center}
\includegraphics[width=7.9cm,height=4cm]{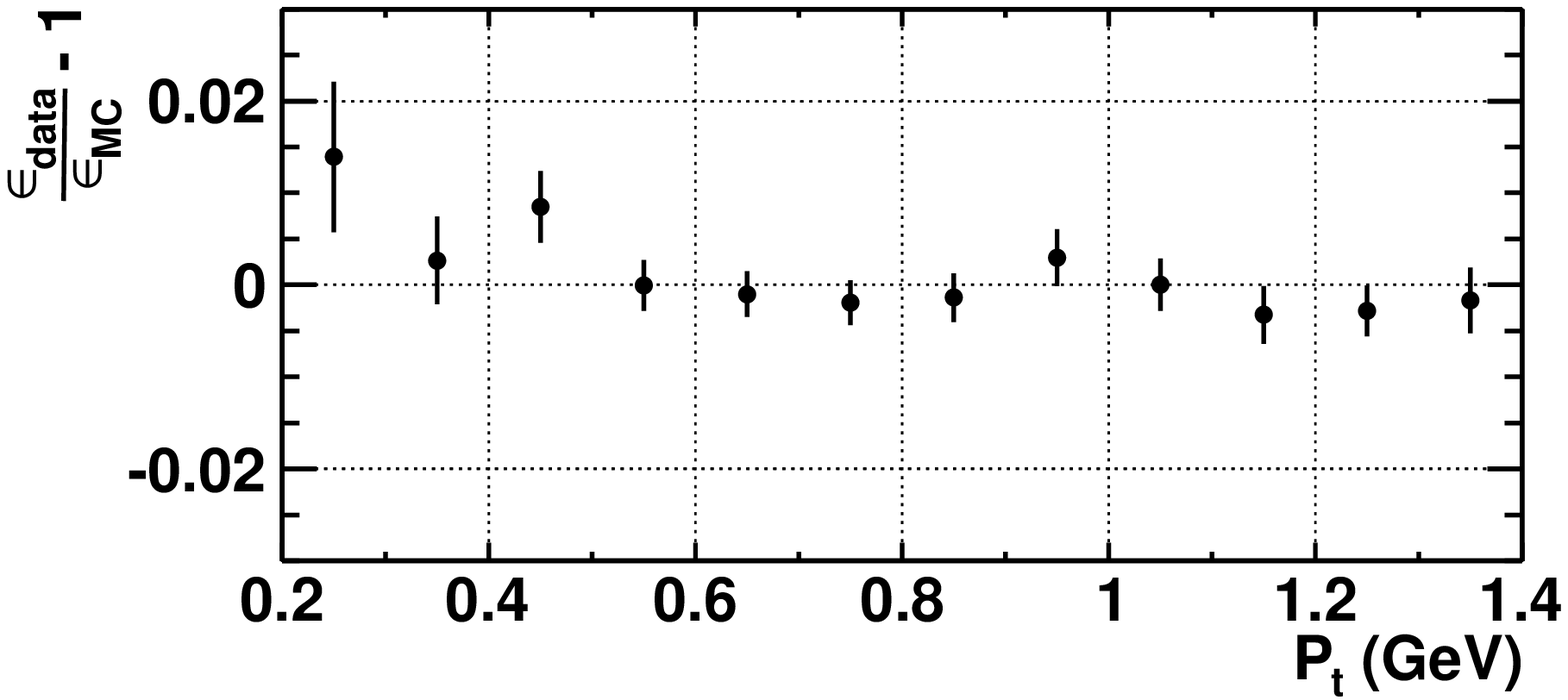}
\centerline{(b)}
\end{center}
\end{minipage}
\figcaption{\label{fig:pieff}   The tracking efficiency of $\pi$ from $J/\psi\rightarrow \pi^{+}\pi^{-}\pi^{0}$. (a) tracking efficiency in different transverse momentum and (b) efficiency difference.  }

\subsection{Momentum resolution}
Protons from $J/\psi\rightarrow p\bar{p}$ are used to check the difference of momentum resolution. Fig.~\ref{fig:pmomentum} shows the momentum distributions of the proton. The resolutions are listed in Table \ref{tab:pmomentum}. The new method has a significant improvement.

\begin{minipage}{4cm}
\begin{center}
\includegraphics[width=4.0cm]{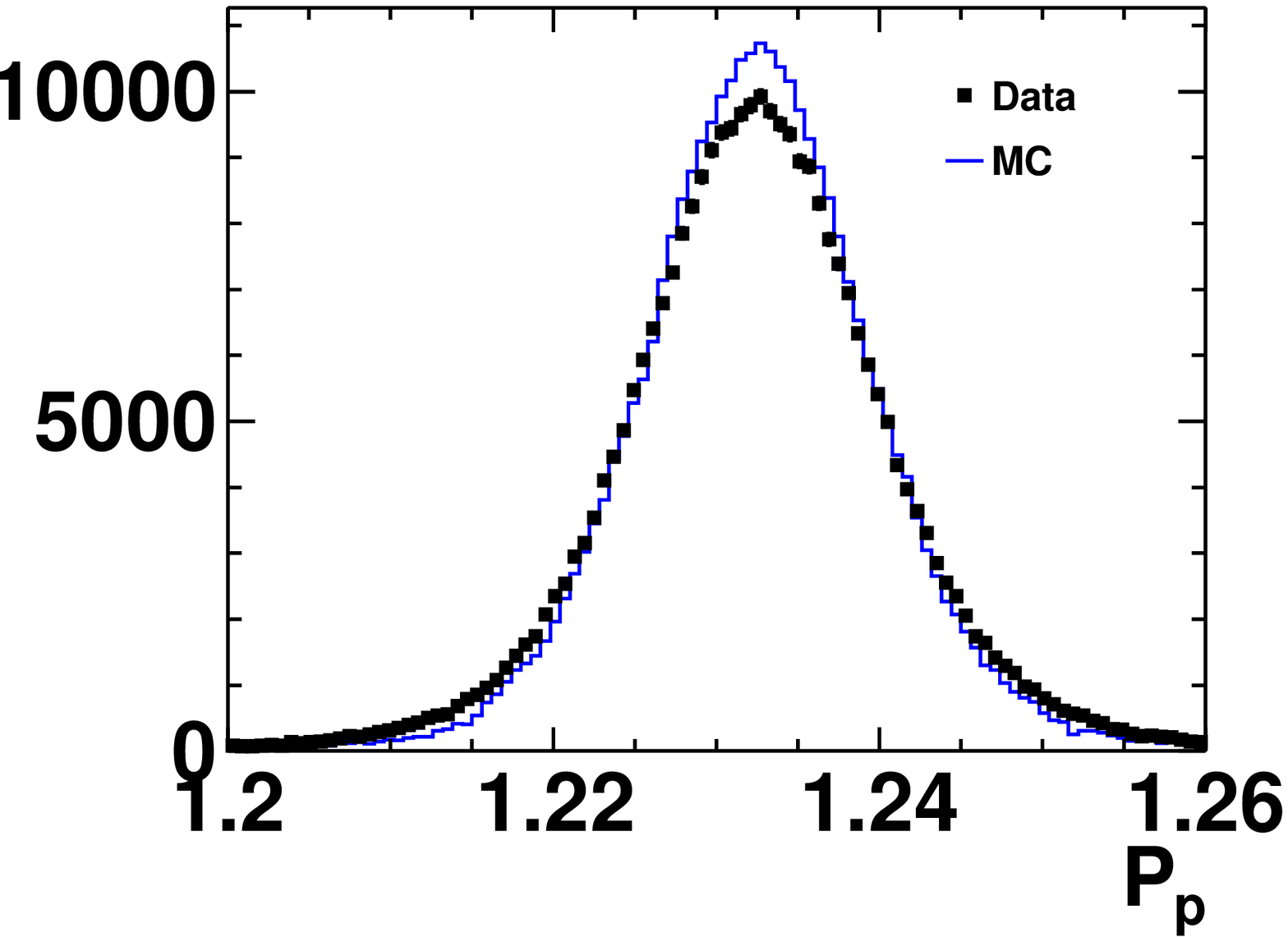}
\centerline{(a)}
\end{center}
\end{minipage}
\begin{minipage}{4cm}
\begin{center}
\includegraphics[width=4.0cm]{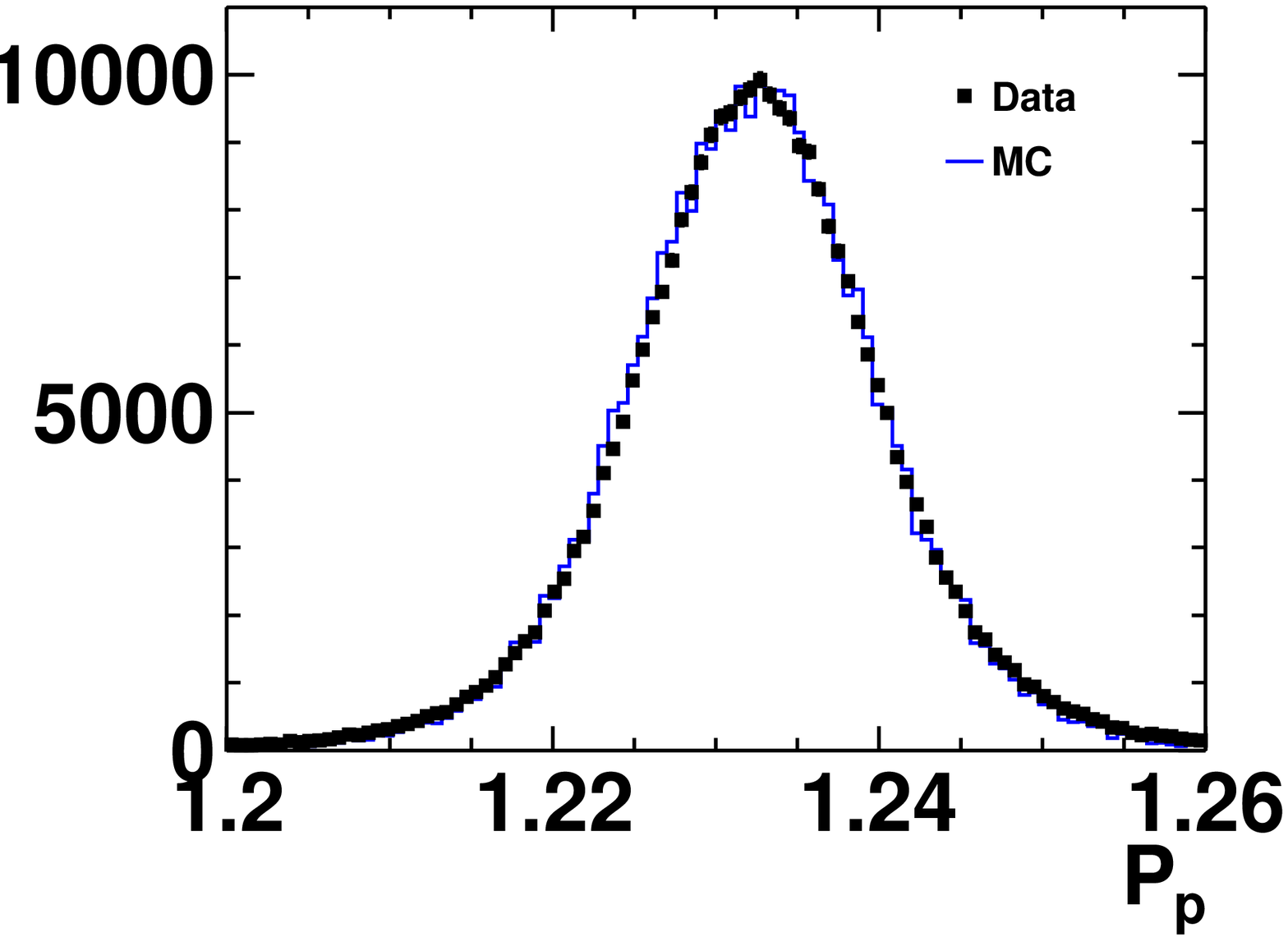}
\centerline{(b)}
\end{center}
\end{minipage}
\figcaption{\label{fig:pmomentum}   The momentum resolutions of proton. (a) previous method and (b) new method. Black dots are data and histograms are MC. }

\begin{center}
\tabcaption{ \label{tab:pmomentum}  Momentum resolutions for data and MC. The difference decreases significantly. }
\footnotesize
\begin{tabular*}{65mm}{c r @{$\pm$} l r @{$\pm$} l}
\toprule
  & \multicolumn{2}{c}{Previous/MeV}   &  \multicolumn{2}{c}{New/MeV}\\
\hline
$\sigma_{\textrm{data}}$     &   7.827&0.013       &  7.827&0.013\\
$\sigma_{\textrm{MC}}$       &   7.097&0.011	   &  7.639&0.035\\
$\frac{\sigma_{\textrm{MC}}}{\sigma_{\textrm{data}}}$-1(\%)  & -9.33&0.21       &  -2.40&0.48\\
\bottomrule
\end{tabular*}
\end{center}

\section{Summary}

In this work, we notice that signals could also cause large residuals. We employ a new category of hits, and revise the model of residual distribution. It can focus on the good-hit efficiency and good hits residual shape directly.

Several advantages of the new method include:
\begin{itemize}
  \item Describing the real data better by splitting MDC hits into good and bad hits.
  \item Reducing the correlation of good hits efficiency and spatial resolution, which is helpful to get better agreements of the tracking efficiency and momentum resolution at the same time.
  \item Accelerating the process of tuning by using an efficient double-Gaussian identifier.
\end{itemize}

Using the new method, the differences in track reconstruction efficiency and momentum resolution decrease significantly. In the future, more effort should be devoted to understand the asymmetry near the sense wire and find out a better model to describe it. Also, the discrepancy of tracking efficiency at low momentum is relatively larger. Thus we should try to improve the agreement for low momentum tracks. If necessary, further improvement can be developed to consider other effects, such as electronic field leakage, $dE/dx$ and so forth.

\vspace{10mm}
\end{multicols}
\vspace{-1mm}
\centerline{\rule{80mm}{0.1pt}}
\vspace{2mm}
\begin{multicols}{2}

\end{multicols}

\clearpage

\end{CJK*}
\end{document}